\documentstyle[11pt,aaspp4]{article}

\slugcomment{Accepted for publication in  Apj Lett., Nov. 21, 1996}

\lefthead{L\'opez-Cruz  et al.}

\righthead{Are Luminous cD Haloes Formed by the Disruption of 
Dwarf Galaxies?}

\begin{document}

\title{Are  Luminous cD Haloes  Formed by the Disruption of 
Dwarf Galaxies?}
\author{Omar L\'opez-Cruz\altaffilmark{1,3,4}, H.~K.~C. Yee\altaffilmark{1,4}, James P. Brown\altaffilmark{1}, Christine Jones\altaffilmark{2}, and William Forman\altaffilmark{2}}

\altaffiltext{1}{Department of Astronomy, University of Toronto, Toronto, M5S 3H8, Canada}
\altaffiltext{2}{Smithsonian Astrophysical Observatory, Cambridge, MA 02138}
\altaffiltext{3}{Also Instituto Nacional de Astrof{\'\i}sica, \'Optica
y Electr\'onica (INAOE), Tonantzintla, Pue., M\'exico} 
\altaffiltext{4}{Visiting Astronomer, Kitt Peak National  Observatory. 
KPNO is operated by AURA, Inc.\ under contract to the National Science
Foundation.} 



\begin{abstract}
From a total sample of 45 Abell clusters observed by the {\em
Einstein} X-ray observatory, we present the results on the galaxy 
luminosity function (LF) for  a group of seven clusters that were
identified by the morphology of their LFs. The LFs were derived using
photometric data to a completeness limit $\sim 5.5$ magnitudes below
$M^{*}$.  We found that a single Schechter function with an average
$\alpha \approx -1.0$ gives a good fit to these individual LFs within
the magnitude range.  These seven clusters have common properties,
which indicate they form a homogeneous class of dynamically evolved
clusters that can be characterized by the presence of a dominant cD
galaxy, high richness, symmetrical single-peaked X-ray
emission, and high gas mass. On the other hand, steep faint-end
slopes ($-2.0 \leq \alpha \leq -1.4$) are usually detected in poorer
clusters. Our result gives a direct indication that the faint-end
slope of the galaxy LF is subject to environmental effects.  We
propose that the {\em flatness} of the faint-end slope in these
clusters results from the disruption of a large fraction of dwarf
galaxies during the early stages of cluster evolution. The stars and
gas from the disrupted galaxies are redistributed throughout the
cluster potential during violent relaxation. This heuristic scenario
can explain the origin of the luminous haloes of cD galaxies and a
large fraction of the gas content in the intracluster medium as a
by-product. The correlation between the cluster gas mass determined
from the modeling of the X-ray emission and the cD halo optical
luminosity is presented to support the proposed model.
\end{abstract}




%

\section{INTRODUCTION}

Twenty-two years ago, Oemler (1974, hereafter
\cite{O74}) in his seminal paper  classified a  sample 
of 15 clusters, based on the morphology of the constituent galaxies,
into three groups: spiral rich, spiral poor, and cD. The luminosity
functions (LFs) for each group indicate a clear progression in the
flattening of the faint-end slope (see Figure 11 in
\cite{O74}), which follows the dynamical sequence 
``from-irregular-to-regular'' that underlies every morphological
classification scheme for clusters of galaxies.  Dressler's (1978)
study showed that some cD clusters have flatter faint-end slopes than
the universal value of $\alpha = -1.25$ introduced by
\cite{Sch76}. Dressler argued that LF of the cD clusters A401 and
A2670, having $\alpha= -1.00$, indicated true departures from
universality generated by environmental effects. However, indications
of universality of the LF were reported in some subsequent studies
(e.g., \cite{Lug86}, \cite{Col88}).

Recently, the galaxy LF has been sampled to very faint levels in
several galaxy clusters, e.g., the nearby Virgo (\cite{San85}) and
Coma (\cite{T&G93}) clusters. In general, there is a good agreement
between the various LF determinations; for example, the value for the
faint-end slope ($\alpha =-1.4$) reported by Thompson \& Gregory
(1993) has been confirmed by \cite{Ber95} in their ultra deep study of
Coma. \cite{Dri94} determined the LF for Abell 963 ($z=0.21$)
reporting a steep faint-end slope ($\alpha = -1.8$). \cite{Dep95} have
reported an extreme value of $\alpha \approx -2.2$, obtained from the
combined galaxy counts of 4 low-redshift clusters. All these studies
imply that there is a marked tendency to detect steep faint-end slopes
($\alpha \leq -1.4$) when dwarf galaxies (dGs) are included. This
seems to agree with the Cold Dark Matter (CDM) scenario (e.g.,
\cite{KWG93}), where the formation of low-mass systems is expected to
be favored.  However, these low-mass systems could be greatly
influenced by their environment (see \cite{F&B94} and
\cite{G&W94}) and some variations should be observed. 

The galactic environment also affects the formation and
properties of the largest galaxies: cD galaxies are regarded as
physically different from ellipticals (\cite{K&D89}) with a different
formation history. cDs are believed to be the product of dynamical
processes which took place during cluster formation (\cite{Mer84}) with
additional growth from galactic cannibalism, cooling flows, or the
accumulation of tidal debris. Although there is no clear evidence to
indicate which of these mechanisms is dominant, the consensus is that
cD galaxies originate by processes solely related to clusters, as no
cD galaxy has been found in the field (\cite{Scm88}).

This {\em Letter} is the first of a series of papers reporting on the
results from a multi-color photometric survey designed to investigate
the optical and X-ray properties in a sample of 45 low redshift
clusters (\cite{Loc96}).  The purpose of the present {\em Letter} is
twofold: first, we report results on the LF for seven clusters (see
Table 1) that have been classified as {\sf flat-LF} (\cite{LCY95});
second, we propose a heuristic model based on the disruption of dGs to
explain the nature of these clusters. Throughout this {\em Letter}, we
assume $H_{\small 0} = 50{~\rm km~ s}^ {-1} {\rm Mpc}^{-1}$ and
$q_{\small 0} = 0$.

\section{OBSERVATIONS}
Properties of the clusters reported here are presented in Table 1.
The observations were carried out at KPNO using the 0.9m telescope and
the T2KA CCD during 1992-93. This array covers a field of view of
$23{\arcmin}
\times 23{\arcmin}$ with a pixel scale  of $0.68\arcsec$
pixel$^{-1}$. Direct images were taken under photometric conditions
using three bands (Kron-Cousins {\sl R, I,} and {\sl B}) covering the
inner regions of the clusters. Image preprocessing was done with {\sl
IRAF}; object finding, star/galaxy classifications, photometry and
generation of catalogs were performed using the photometry code {\sl
PPP} (\cite{Yee91}). On each field, $\gtrsim 3000$ objects were
identified and classified.  To avoid crowding in the central
regions, the brightest cluster members (BCMs) were removed by modeling
their isophotes (\cite{Bro96}) before executing the photometry for
the rest of the galaxies. Standard stars from \cite{Lan92} were
observed nightly to derive the transformation to the
standard Kron-Cousins system.  The average $5 \sigma$ detection
magnitude limit reached in each band is 23.5 in {\sl B}, 22.5 in {\sl
R}, and 21.5 in {\sl I}; the 100\% completeness limit is $\sim 0.8$
mag brighter.  In addition, five high galactic latitude random
control-fields were observed to determine field galaxy counts. The
latter  observations give us the ability to generate field-corrected
LFs, along with the uncertainties due to field-to-field variations.

\section{GALAXY LUMINOSITY FUNCTION DETERMINATIONS}        

Color information is used to discriminate galaxies whose colors are
too red to be considered as clusters members, i.e., galaxies much
redder than the color-magnitude relation for early type galaxies at
the cluster's redshift are omitted in the generation of the LFs (see
Table 1). The LFs are generated for the remaining galaxies using the
statistical approach developed in
\cite{O74}. First, the observed apparent magnitudes of the galaxies
are converted to absolute magnitudes using the cluster redshift. These
magnitudes are corrected for galactic absorption (\cite{Bur82}) and
for K--dimming (\cite{Coh80}).  Next, the galaxy counts are binned by
absolute magnitude into 0.5-mag bins. To derive the 
field-corrected LFs, the same color-cut applied to the cluster in
question is applied to the control fields before the field counts are
derived. The generated field galaxy counts are subtracted bin by bin
from the cluster galaxy counts.

A Schechter function expressed in magnitudes (cf. \cite{Col88}) is
used to model the generated LFs:
$n_{S}(M)dM={\mathcal{K}}N^{*}\exp{\left\{{\mathcal{K}}(\alpha+1)(M^{*}-M)-
\exp{[{\mathcal{K}}(M^{*}-M)]}\right\}}dM$,
where $M^{*}$ is the characteristic magnitude, $\alpha$ is effectively
the faint-end slope, and ${\mathcal{K}}= \frac{\ln{(10)}}{2.5}$. The
parameters $\alpha$ and $M^{*}$ are obtained by minimizing the
$\chi^{2}$ statistic which depends on the model and the observed LF.
The expected number of galaxies per magnitude bin corrected for the
 finite bin size $\Delta M$ is given by (Schechter 1976):
$N_{ei}=n_{S}(M_{i}){\Delta}M + \left. \ {\frac{d^{2}n_{S}}{dM^{2}}}\right|_{M_{i}}
\frac{{\Delta}M^{3}}{24}$,
with an uncertainty $\sigma= (N_{i}+ 1.69N_{bi})^{\scriptsize
\case{1}{2}}$, where $\sqrt{N_{i}}$ is the estimated Poisson error for
the observed cluster galaxies corrected for the background, and
$1.3\sqrt{N_{bi}}$ is the field-to-field variation per magnitude bin
determined from the control fields.  Following \cite{Dre78}, $N^{*}$
is derived by requiring that the predicted total number of galaxies be
equal to the total number in the observed LF. To compare the LFs of
clusters at different redshifts in a consistent manner and reduce the
effects of superpositions by foreground or background clusters or
groups, the LF are generated for galaxies in a circular area of
$1.5{\rm Mpc}$ in diameter centered on the position of the BCM. The
results are given in Table 1. The LFs and their parametrizations are
depicted in Plate 1. It is apparent that these are relatively good
fits ($\frac{\chi^2}{\nu}\approx 1.0$).

From the 39 of clusters in our sample with $z < 0.14$, where our
observations reached completeness magnitude limits that allow us to
detect unambiguously the presence of steep faint-ends, only the seven
clusters presented here can be adequately characterized with a single
Schechter function. The rest require the combination of two fitting
functions.  As examples, we present the LFs of the poor cluster A1569
and the Coma cluster (A1656). The use of the sum of two Schechter
functions is justified because a single fitting function cannot
account for the significant steady rise in the galaxy counts below
$M_{R} \approx -19.50$. The fitting proceeds by holding ${\alpha_{1} =
-1.0}$ fixed (this $\alpha$ resulted after combining the 45 clusters in 
our sample, the Schechter fit to the combined LF for $M_{R}
\leq -20.0$ gave  $\alpha = -1.04 \pm 0.05$ and $M^{*} = -22.53 \pm
0.09$), and applying the constraint $N^{*}_{2}= 2N^{*}_{1}$ (when
$N^{*}_{1}$ and $N^{*}_{2}$ are free to vary, the resulting fit gave
$N^{*}_{2}
\approx 2N^{*}_{1}$ and the goodness of the fit was similar) to derive the remaining
parameters ($M^{*}_{1},N^{*}_{1},\alpha_2, M^{*}_{2}, N^{*}_{2}$).
The results are shown in Table 2. A consistent trend observed in our
sample indicates that steep faint-ends are detected in poorer
clusters; flatter faint-end slopes are exhibited on average by richer
clusters (\cite{LCY96}), while {\sf flat-LF} clusters, being in
general the most massive, represent an extreme situation. However,
Rood-Sastry binary clusters, appear to deviate from this trend: some
binary clusters are very rich clusters, but their faint-end slopes are
steep. This could be the result of recent cluster-cluster or
cluster-group mergers (\cite{STr90}); e.g., the relatively rich Coma
cluster, the possible the result of the recent merger of at least two
comparable clusters (\cite{Whi93}, \cite{Vik94}). We remark that our
LF parameters for Coma agree very well with the deep studies of
Thompson \& Gregory (1993), Bernstein et al. (1995), and
\cite{S&H96}.
 
\section{DISCUSSION}
The fact that the steepening of the galaxy LF towards faint magnitudes
is not present in all clusters can be taken as evidence against the
{\em universality} of the galaxy LF in clusters.  Our results strongly
support Oemler's classification scheme (\cite{O74}). These seven {\sf
flat-LF} clusters have common properties which indicate they form a
homogeneous class of dynamically evolved clusters that is
characterized by  high cluster richness, the presence of a dominant cD
galaxy (Bautz-Morgan type I, I-II, or Rood-Sastry class cD),
symmetrical single-peaked X-ray emissions with the cD galaxy located
on the X-ray peak, and high gas masses. There have been many attempts
to detect LF variations in galaxy clusters which could be indicative
of the dynamical processes that generate cD galaxies; so far, the
results have been controversial. The sample of {\sf flat-LF} clusters
reported here represent a class where such effects are readily
distinguishable, indicating that dGs are sensitive to environmental
effects.

To interpret the nature of {\sf flat-LF} clusters, we hypothesize that
a large fraction of the dG population inside clusters has been
disrupted during the early stages of cluster formation. The stars from
the disrupted galaxies are redistributed during violent relaxation
throughout the cluster's potential  generating the cD's
halo. Moreover, as a by-product, the gas originally confined in dGs is
incorporated into the cluster and can contribute significantly to the
mass of the intracluster medium (ICM).

To investigate whether the disruption of dGs in clusters is able to
generate the luminosity of cD haloes, we neglect the effects of stellar
evolution and  proceed by postulating  that the total luminosity
from the initial LF is equal to the total luminosity in the
present-day LF plus the luminosity of the cD halo, i.e., ${\mathcal
L}_{init} = {\mathcal L}_{cl} + {\mathcal L}_{cD~halo} $. We find that
the LFs of non{\sf flat-LF} clusters are adequately fit by the sum
of two Schechter functions; therefore, we suggest a similar
parametrization for the initial LFs. Hence, ${\mathcal L}_{init}={
\int_{-28}^{-11}{\mathcal L}\:[n_{S}(M;\alpha_{1}^{\prime},
{M^{\prime}_{1}}^{*}, {N^{\prime}_{1}}^{*})+
n_{S}(M;\alpha_{2}^{\prime},{M^{\prime}_{2}}^{*},
{N^{\prime}_{2}}^{*})] dM}$, where $\mathcal L$ is the galaxy
luminosity in the {\em R} filter, and the limits of integration cover
the range in absolute magnitudes from giant to dwarf galaxies.  We
assume that during cluster formation the giant galaxy population is
not severely affected; therefore, the first term of the initial LF
should be similar to the presently observed LF. Hence, ${\mathcal
L}_{cD~halo}
\simeq \int_{-28}^{-11}{\mathcal L}\:n_{S}(M,\alpha_{2}^{\prime},
{M^{\prime}_{2}}^{*}, {N^{\prime}_{2}}^{*})dM$. We propose a {\em
fiducial} LF for the second term of the initial LF with a relatively
steep faint-end slope $\mathbf{\alpha_{2}^{\prime} = -1.8}$, and the
average characteristic magnitude for non{\sf flat-LF} clusters,
${\mathbf{{\mathit{M}}^{\prime}_{2}}^{*}= -19.0}$. It follows, that
the normalization parameter ${N_{2}^{\prime}}^{*}$ in the initial LF
for dGs can be estimated using ${\mathcal L}_{cD~halo}$.  We have
modeled the surface brightness distribution of the cD galaxies, the
details about the modeling are given elsewhere
(\cite{Bro96}).  Only three clusters (A1795, A2420, A2670) obey
Schombert's (1988) definition of cD halo, i.e., the presence of the cD
haloes is indicated by an inflection in the outer surface brightness
galaxy profile. Nevertheless, because the BCMs in A399, A401, A1650,
and A2029 are very luminous galaxies and the slopes of their surface
brightness profiles are much shallower than normal early-type
galaxies;  we suggest that these BCMs are dominated by their
haloes. We propose the term halo-dominated cD galaxy to classify this
kind of BCM. In Table 1 we present the derived values of
${N_{2}^{\prime}}^{*}$ that fulfill the conditions required by our
model, the resulting
$\langle\frac{{N_{2}^{\prime}}^{*}}{{N_{1}^{\prime}}^{*}}
\rangle =1.8$ is similar  to that determined from 
fitting non{\sf flat-LF} clusters. We have shown that using reasonable
Schechter function parameters, we are able to account for the light in
cD haloes as arising from the disruption of dGs.

Next, the derived parameters for the initial LF are used to test
whether the gas from the disrupted dGs is sufficient to account for the
ICM mass in the central cluster regions. We note that \cite{Tre94}
suggested a similar origin for the ICM, with the gas being
incorporated into the ICM via supernova driven winds. However, more
detailed studies have shown that supernova driven winds are
insufficient to incorporate the large amounts of gas that have been
measured in the ICM (\cite{N&C95},
\cite{Gib96}).  Our model gives an efficient mechanism to incorporate
the gas from dGs into the ICM because it relies on the total
disruption of the dGs. We can repeat the trentham (1994) analysis for
each of our cluster to estimate the amount of gas originally confined
in dGs. The amount of gas that can be removed from dGs is
given as a fraction $\gamma$ of their initial total mass. Since we
expect the dGs to be gas rich, we choose $\gamma = 0.33$ (Trentham
1994).  The supernova driven wind models can only allow much lower
values of $\gamma \leq 0.09$. We assume a mass-to-light ratio of the
form ${\mathcal \frac{M}{L}
\propto {\mathcal L}^{\beta}}$ (\cite{K&D89}).  Hence, the total gas
mass contributed by the disrupted galaxies, ${\mathcal M}_{gas}$, is
given by (cf. \cite{N&C95}):
\begin{equation}
{\mathcal{M}}_{gas} =
{N^{\prime}_{2}}^{*}{\mathcal{M^{*}}}\frac{\gamma}
{1-\gamma}\int_{\mathcal{M_{-}}}^{\mathcal{M_{+}}}\exp{\left[-\left(
{\mathcal{\frac{M}{M^{*}}}}\right)^{\frac{1}{\beta + 1}}\right]}
\left({\mathcal \frac{M}{M^{*}}}\right)^{\frac{\alpha_{\tiny 2}^{\tiny \prime} + 1}{\beta + 1}}
\frac{d{\mathcal{M}}}{(1 + \beta){\mathcal{M^{*}}}}.
\end{equation}
The limits of integration are ${\mathcal M}_{-} = 1\times
10^{4}{\mathcal M_{\tiny \sun}}$ and ${\mathcal M}_{+} = 1\times
10^{11}{\mathcal M_{\tiny \sun}}$, and the mass-to-light ratio for a
${M_{2}^{\prime}}^{*}$ galaxy is given by
${\mathcal{\frac{M^{*}}{L^{*}}}} \approx 10$ in the {\em R}
filter. The only free parameter in Equation (1) is $\beta$. To derive
$\beta$ we set ${\mathcal M}_{gas} = {\mathcal M}_{X}$, where
${\mathcal M}_{X}$ is the observed gas mass inside the inner 1.5 Mpc
deduced from the modeling the X-ray surface brightness distribution,
observed with {\em Einstein-IPC} (\cite{J&F96}).  The resulting range
$-0.30\leq \beta\leq -0.24$~ (see Table 1) is considerably narrower
than the one derived by Trentham ($-0.42 < \beta < -0.27$). Our
$\langle\beta\rangle = -0.28$ is consistent with the observationally
derived $\beta = -0.22 \pm 0.09$ (\cite{Kor90}) for dwarf spheroidal
galaxies (dSphs). We note that there is evidence that dSphs are
affected by the cluster's environment more severely than dwarf
ellipticals (\cite{T&G93}).  Hence, we suggest that the disrupted
galaxy population is dominated by dSphs. Moreover, the resulting
slope, $\frac{\alpha - \beta}{\beta + 1} \approx -2.1$, in the initial
mass function roughly agrees with the $\sim -2$ slope predicted by
CDM. Based on the close match between derived parameters and their
observed counterparts, we conclude that dGs can be considered the main
contributor to the intracluster light and the gas in the ICM. A more
comprehensive picture requires the contribution of the giant galaxies
to supply the core of the cD galaxy ({\cite{Mer84}) and metals for the
enrichment of the ICM (\cite{N&C95}, \cite{Gib96}).

If cD haloes and the gas in the ICM have a common origin, then, a
correlation between the luminosity of the cD halo and the gas mass of
the ICM is expected. Figure 2 shows a clear correlation between the
luminosity of the cD halo and the cluster's gas mass, a similar
correlation was reported  by (\cite{Scm88}). A simple linear
regression analysis gives ${\mathcal M}_{X}
\propto {\mathcal L}_ {cD~halo}^{0.64 \pm 0.16}$ with a correlation
coefficient ${\rm R}=0.87$.  We have also plotted the cD halo
luminosity of NGC 4874 (Coma's BCM). If this galaxy is included with
estimated gas mass for the whole cluster; the significance of the
correlation diminishes (${\rm R}=0.70$). However, if the average mass
of the isolated X-ray component around NGC 4874 (\cite{Vik94}) is
considered, then NGC 4874 falls on  the line derived for the cDs
in flat-LF clusters (see Figure 2). This result  reaffirms the
cluster-cluster merger nature of Coma.

No mechanism for galaxy disruption or merging would be effective
in the present-day conditions in clusters (\cite{Mer84},
\cite{STr90}). Thus, it has been suggested  that the process that gave
rise to the cDs took place during an early stage of cluster
evolution. A plausible mechanism is that clusters are assembled by
merging groups of galaxies (\cite{STr90}). Processes like collisional
tidal striping (\cite{Agu85}) can play a major role in disrupting dGs
in groups of galaxies because groups have smaller velocity dispersions
than virialized clusters. During the initial stages of cluster
evolution, the cluster's varying tidal field could be considered an
important agent to limit the size of galaxies. Giant galaxies and dGs
are both exposed to the effects of the tidal field; however, very
massive compact objects such as giant ellipticals are not severely
affected (\cite{Mer84}).

We have suggested a case to explain the origin of {\sf flat-LF}
clusters based on the disruption of dGs. This process results in the
formation of the luminous haloes of cD galaxies and contributes
significantly to the gas in the ICM. Other observed properties
indicate that {\sf flat-LF} clusters are a homogenous class of
dynamically evolved clusters. Whether all cD galaxies are found in
{\sf flat-Lf} clusters is a difficult question. There are other
effects that can mask the flatness of the LF faint-end: Coma, for
instance, has a cD galaxy, but it cannot be classified as a {\sf
flat-LF}; perhaps the result of  Coma's   complex
cluster-cluster merger history. Nevertheless, evidence of dG
disruption is found in Rood-Sastry  cD clusters as indicated
by their LF parameter {$\alpha_{2} \geq -1.4$} (\cite{LCY96}).


\acknowledgements
We thank KPNO for the use of their superb facilities and the staff for
their prompt assistance. The authors are very much indebted to
G. Djorgovski, D. Merritt, A. Oemler, M.F. Struble, M.J. West, R.
Kirshner, S. Tremaine, R. Carlberg, N. Kaiser, S. Lilly, J. Huchra,
B. McNamara, L. David, J. Mohr, D.J. Eisenstein, L.F. Barrientos,
R. van de Weygaert, and P. Papadopoulos for very useful conversations
and comments. OLC acknowledges the support from CONACYT-M\'exico,
INAOE, NASA grant NAG8-1881, NASA contract NASA-39073, and the
Smithsonian Institution. This research is supported, in part, by a
NSERC operating grant to HKCY.

\clearpage

\clearpage

\begin{figure}
\plotone{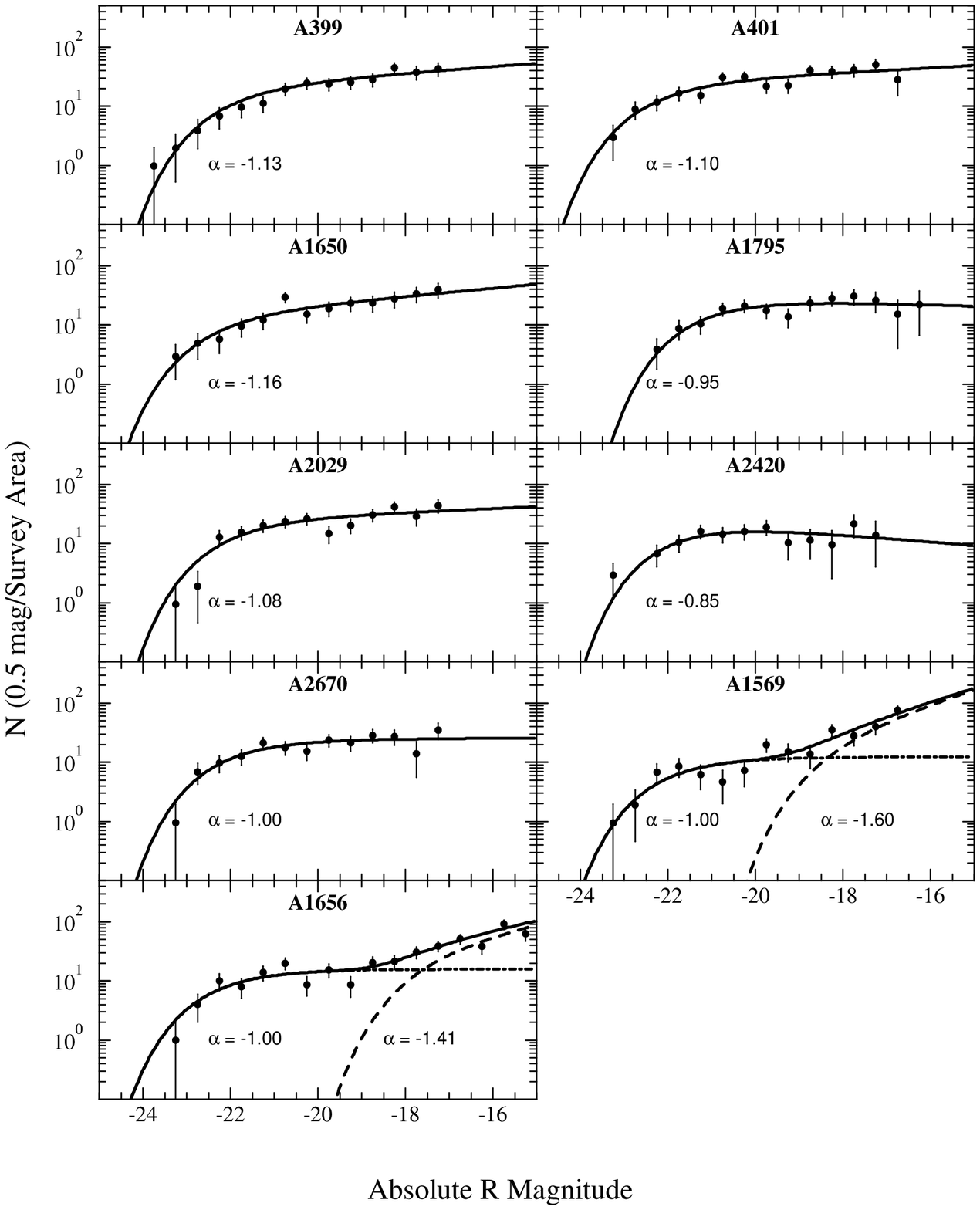}
\caption{Luminosity functions for  {\sf flat-LF} clusters. The values of
$\alpha$ are indicated, cD galaxies were not included in the
fitting. The clusters A1569 and A1656 are presented for
comparison. The sampled cluster diameter is 1.5 Mpc, except for A1656
(${\mathsf \Phi} \approx 0.9\; {\rm Mpc}$)}
\end{figure}
\clearpage

\begin{figure}
\plotfiddle{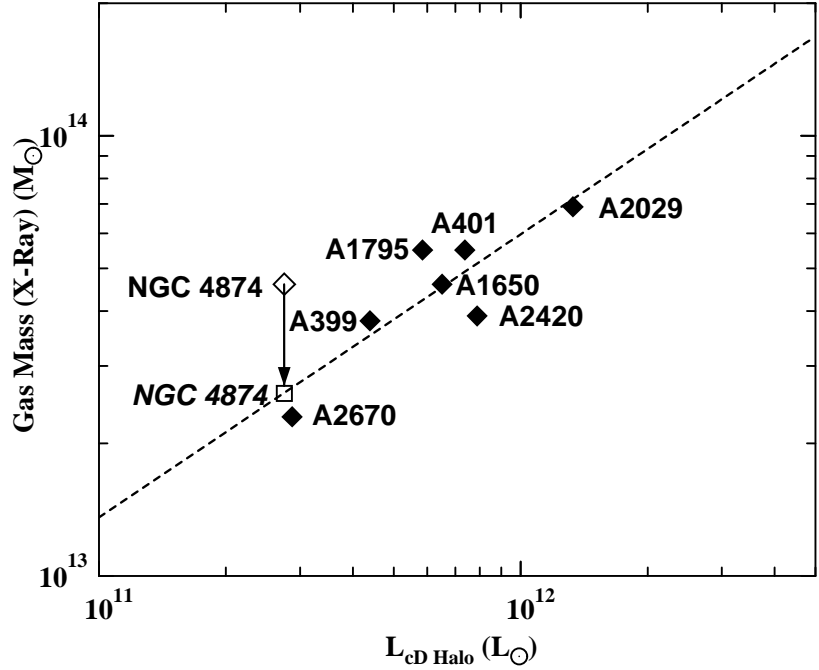}{5cm}{-90}{50}{50}{-190}{300}
\caption{Gas Mass (units of ${\mathcal M}_{\tiny \sun}$) v.s. optical 
luminosity (units of $L_{\tiny \sun}$) of the cD halo, the dashed line
is a least-squares fit (NGC 4874 was not included) that gives
${\mathcal M}_{X} \propto L_{cD~halo}^{0.64\pm 0.16}$. The square
indicates the X-ray mass determined by Vikhlinin et al. (1994).  This
correlation indicates that the cD halo and the gas in the ICM may have
a common origin.}
\end{figure}

\end{document}